# Heat spreader with parallel microchannel configurations employing nanofluids for near–active cooling of MEMS

**Lakshmi Sirisha Maganti**[1, a], **Purbarun Dhar**[2, b], **T Sundararajan**[1, c, *] and **Sarit Kumar Das**[1, d, *]

[1] Department of Mechanical Engineering, Indian Institute of Technology Madras, Chennai–600036, India

[2] Department of Mechanical Engineering, Indian Institute of Technology Ropar, Rupnagar–140001, India

[a] E-mail: lakshmisirisha.maganti@gmail.com
[b] E-mail: purbarun@iitrpr.ac.in
[*, c] Corresponding author: E-mail: tsundar@iitm.ac.in
[*, d] Corresponding author: E-mail: skdas@iitrpr.ac.in
Phone: +91-1881-24-2100

## Abstract

While parallel microchannel based cooling systems (PMCS) have been around for quite a period of time, employing the same and incorporating them for near–active cooling of microelectronic devices is yet to be implemented and the implications of the same on thermal mitigation to be understood. The present article focusses on a specific design of the PMCS such that it can be implemented at ease on the heat spreader of a modern microprocessor to obtain near-active cooling. Extensive experimental and numerical studies have been carried out to comprehend the same and three different flow configurations (U, I and Z) of PMCS have been adopted for the present investigations. Additional to focussing on the thermofluidics due to flow configuration, nanofluids (as superior heat transfer fluids) have also been employed to achieve the desired essentials of mitigation of overshoot temperatures and improving uniformity of cooling. Two modelling methods, Discrete Phase Modelling

(DPM) and Effective Property Modelling (EPM) have been employed for numerical study to model nanofluids as working fluid in micro flow paths and the DPM predictions have been observed to match accurately with experiments. To quantify the thermal performance of PMCS, an appropriate *Figure of Merit* (FoM) has been proposed. From the FoM It has been perceived that the Z configuration employing nanofluid is the best suitable solutions for uniform thermal loads to achieve uniform cooling as well as reducing maximum temperature produced with in the device. The present results are very promising and viable approach for futuristic thermal mitigation of microprocessor systems.

## 1. Introduction

The advancement of civilization increases requirement for smaller, faster and powerful electronic devices and mechanized components. Decrease of size conglomerated with increment in power output results in enhanced high heat flux generation because of billions of microelectromechanical components working at high frequency. Often, the heat flux generated by such modern devices is comparable to heat flux generated by conventional thermofluidic devices such as jet engines. Increase in heat generation rates from such sophisticated MEMS demands smart cooling techniques. Developing such thermal management systems has become more challenging to thermofluidic industries and academicians. There have been many efforts undertaken by the research community to overcome the problem of high heat dissipation from small areas. It is Tuckerman & Pease [1] who first came out with a ground-breaking idea of generating high heat transfer coefficients within heat exchangers by introducing the concept of microchannels (high surface area to volume ratio). Over the years the design of microchannel cooling systems has experienced advances towards perfection in order to achieve higher thermal efficiency. As part of this endeavour, there was a time when the research community concentrated on fundamental aspects, such as the applicability of the Navior–Stokes equations to predict the hydrodynamic characteristics in micro flow paths. Qu and Mudawar [2] reported both numerically and experimentally that conventional Navior-Stokes equation can be employed for accurate prediction of flow characteristics in micro flow paths. In the beginning there was a conflict of applicability of Navior-Stokes equations for micro flow paths but that was cleared from the reported studies [3-6]. The disagreements are mainly caused due to measurement inaccuracy, defectiveness in the test section due to fabrication challenges, entrance effects and surface roughness.

Having understood fundamental accepts in micro flow paths, researchers and academicians have concentrated on the design aspects of parallel microchannel cooling systems (PMCS) to obtain high thermal efficiency micro heat exchangers. The challenges to be addressed for such complex flows to achieve high thermal performance in cooling microelectronic devices have been extensively reviewed [7, 8] and suggestions have been proposed. Attempts have been made to check the applicability of existing theoretical models of mini and macro flows to micro flows and it has been concluded that existing theoretical models are ineffective to predict flow characteristics of micro flows [9]. The most important occurrence which can hamper the thermal performance of PMCS is non uniform distribution of fluid among parallel channels; also termed as flow maldistribution. The effect of flow maldistribution on thermal performance of parallel channel micro heat exchanges have been studied [10-12] and concluded that fluid maldistribution will induce non-uniform cooling of device which leads to formation of unintended hot spots. There have been lot of efforts made by research groups to understand flow maldistribution among parallel microchannels [13-16] and several suggestions to reduce flow maldistribution have been proposed. The major objectives of PMCS are reducing maximum hot spot temperature and increasing the uniformity of cooling so as to obtain higher thermal safety of electronic devices. One of the best methods to achieve enhanced cooling is employing high thermal efficiency coolants such as nanofluids [17-19]. Nanofluids have been reported to behave like smart fluids at different temperatures [20]. Such smart effects are a direct consequence of particle migration effects such as Brownian motion, thermophoresis, Saffman lift, drag etc. and hence cannot be modelled as homogeneous effective property fluids. The correct methodology to model such fluids has been reported as the discrete phase model (DPM) [21, 22] and researchers [23, 24] have modelled nanofluid as non-homogeneous fluids and observed good agreements with experiments.

The cooling mechanisms used in modern day microprocessors, such as heat sinks with fans, heat pipes, etc. are expected to fall short of the cooling requirements in the near future due to increasing microchip calibre which leads to increased heat loads. Essentially, active cooling systems such as on chip cooling by microchannels are required. However, on-chip cooling is not at all a feasible option as the current design of microprocessors needs to be redesigned to achieve this. However, complete eradication of the auxiliary cooling systems like heat sinks, heat pipes is possible by a near active cooling system where the microchannel system is fabricated onto the heat spreader integrated to the microprocessors. It essentially serves the two fold purpose of removing additional thermal resistances from the system as well as dissipates the heat through closest possible approach to on-chip cooling. The concept design of the same has been illustrated in figure 1. The present article focuses on such a cooling system, both experimentally and computationally. A comprehensive study on the usage of various PMCS flow domains and their effectiveness to mitigate flow maldistribution induced hot spots has been presented. Nanofluids have been employed as advanced coolants to counter the thermal challenges where normal fluids like water fail to provide satisfactory performance. Overall, the findings of the present paper provide a concise overview of the potential the discussed method holds for cooling futuristic microdevices.

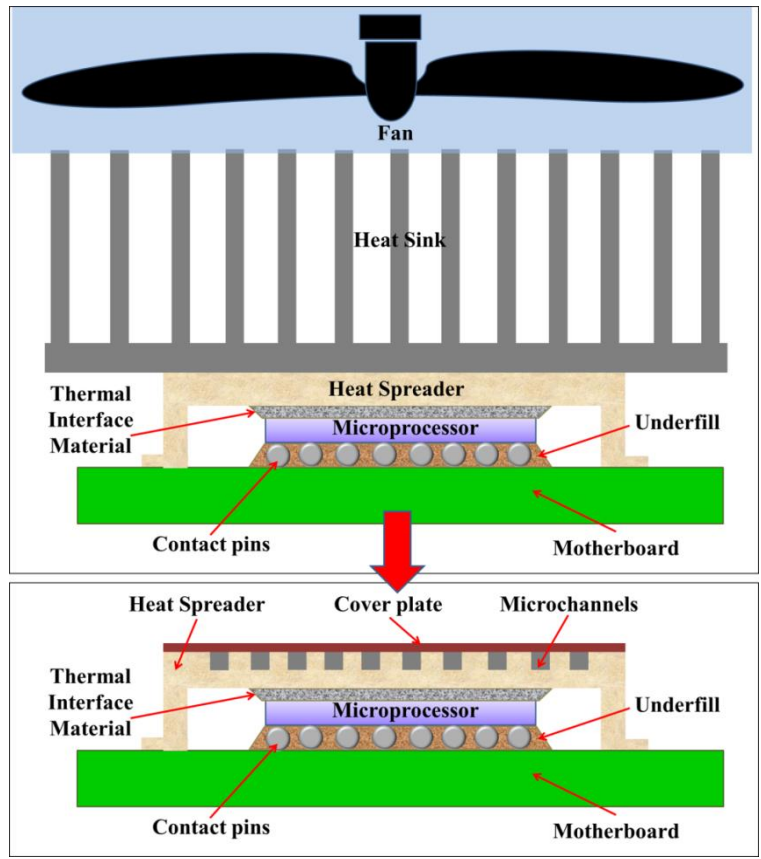

**Figure 1:** (Top) Conventional thermal management technique for present day microprocessor systems. (Bottom) Proposed microprocessor cooling system employing microchannel based heat spreader, thus eliminating heat sink and fans as well as not involving the technologically involved concept of microchannels on the silicon microprocessor itself.

## 2. Materials and methodology

### 2.1. Experimental details

The test section used in the present study has been fabricated so as to mimic a heat spreader system of a microprocessor assembly, as conceptualized in Fig. 1. Accordingly, the test section has been fabricated using aluminium blanks (grade 6061) using a CNC micromachining workbench. Initially, blanks of aluminium of thickness 5 mm are cut 60 mm x 60 mm in size. The size has been selected keeping in mind the size of the heat spreader of a general Intel core i7 microprocessor. The additional region has been kept for ease of closing the channels with a top panel. In the cut blank, a groove 50 mm x 50 mm is machined using a micro end mill cutter in a CNC micromachining facility. The channel assembly will be machined within the grooved region and the groove also ensures proper sealing of the channels post machining. The thickness of aluminium within the grooved region is kept 2 mm to facilitate minimum temperature drop across the metal. The internal base of the groove has been surface machined to mirror finish to ensure minimal friction factor for the flows as well as ensure leak proof bonding of the channels to prevent flooding from the test section. Further, the manifold and channel system has been machined employing micro end mill cutters. Post machining, the dimensions of the channels have been cross verified employing a tool makers microscope and the surface roughness of the channel base has been measured with a surface profilometer (illustrated in fig. 2).

The microscopy reveals that the machining process has ensured channels with an accuracy of ± 5 % with respect to the hydraulic diameter. The surface roughness measurment shows that the average roughness value ranges from within 0.5–1μm and hence the channel system can be considered as smooth. Post fabrication, the top portion of the channel system has been sealed using an acrylic sheet, machined to size so as to perfectly fit the grooved region. The acrylic is press fit into the groove , thus ensuring mechanical bonding and closing of the channels. The regions of the acrylic sheet protruding above the groove after fitting is sealed to the aluminium blank employing Silicone RTV, further ensuring bonding. In order to pump fluid into the system and measure pressure drop data, microholes have been machined in the acrylic using a microdrill. The drilled holes contain two regions along the axis of the hole. Inititally, a microhole is machined with width equal to the manifold (for sending and collecting fluid) or the channel width (for pressure tappings). Then, a larger concentric region is machined to a small depth so as to fix the piping. Polyurethane pipes are employed to ensure that there is no thermal damage to the piping network during heating and these are bonded to the test section using Silicone RTV. K type thermocouples have been inserted through micro holes at strategic locations to note the temperatures of the domain during the experiments. The specific locations have been identified from computations and the thermocouples accordingly positioned. The bead of the thermocouple is covered in thermal paste before insertion into the holes to ensure accurate reading of the local temperature.

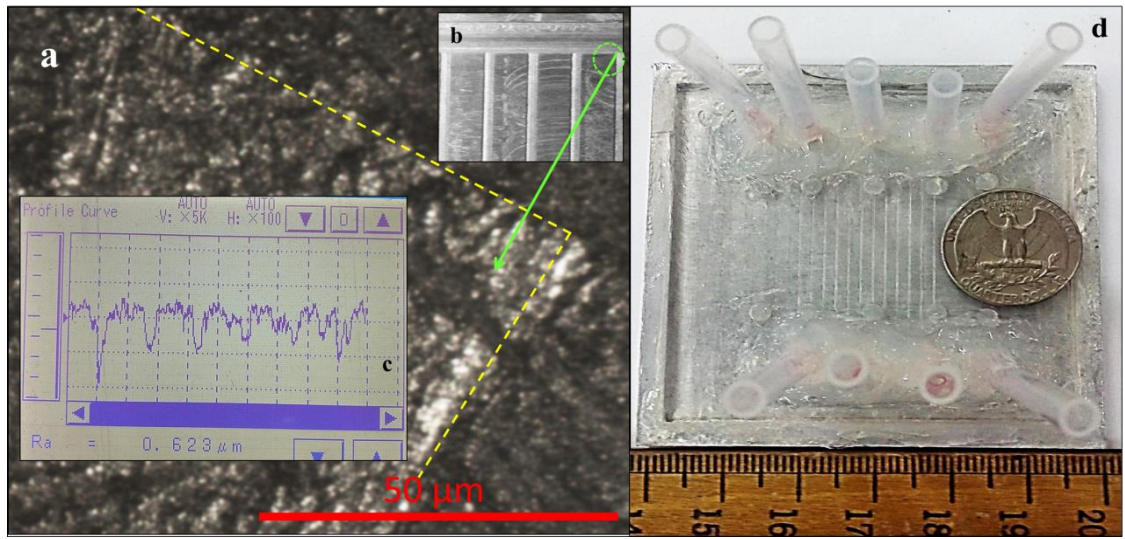

**Figure 2: (a)** Assessment of channel dimensions post machining through microscopy **(b)** Illustration of the PMCS post machining **(c)** Representative channel roughness profile **(d)** Illustration of the complete PMCS after bonding and with flow paths and pressure taps. A wooden ruler and a USA quarter dollar coin has been used for scale.

For the present experimental studies, an in–house test rig has been setup by assembling components as required. The schematic of the complete test rig has been illustrated in fig. 3. In precis, the experimental rig consists of a constant temperature bath (F25-MA, Julabo GmbH, temperature stability ~ ±0.05 K) which is employed to house the fluid sample as well as maintain its temperature to the requisite inlet temperature to the test section. The fluid is withdrawn and pumped into the test section from the thermal bath employing a dual acting syringe pump set (Dual-NE-1000, New Era Pump Systems, USA). In order to ensure minimal temperature drop for the fluid from the thermal bath to the test section, the flexible piping used to channel the fluid to the test section is insulated using foam pipe insulators. To ensure that there is continuous infusion by the dual acting pump and that no air bubbles are introduced between the switching operation of the infusion and the withdrawing pumps, the whole piping circuit is first filled with the working fluid using a gear pump (Reglo-Z, Ismatec, Germany). Also, as the continuous flow due to syringe pump operation becomes hindered towards higher Reynolds numbers, the gear pump has been employed for Re > 50. The pressure drop across the channels within the test section are measured using a differential pressure transducer (PS-100PSID-D, Alicat Scientific Inc., US). Anti corrosive version of the pressure transducer has been selected in order to ensure proper working in case of nanofluids as well. The heating arrangement for the present work has been designed and fabricated in-house. The arrangment consists of a laminated heating panel equal in size as that of the test section. The heating panel is supplied power using a controller panel and the input power supplied to the test section is obtained using a multimeter. The heat input to the test section is restricted to required limits employing current control within the controller unit. Temperatures within the test section at requisite regions are measured using a set of calibrated K type thermocouples (accuracy ±0.5 $^{o}$C). The thermocouples are housed within the assembly using micro drill holes in the test section. To ensure safety limits for the temperature at the hot spot regions (the location as determined from simulations), a PID based temperature controller unit (Selec TC–513, India) is also used to cut off power to the heater in case the hot-spot temperature overshoots 90 $^{o}$C. The test section is mounted onto the heater assembly firmly using screws and thermal paste is applied in between to reduce contact resistance.

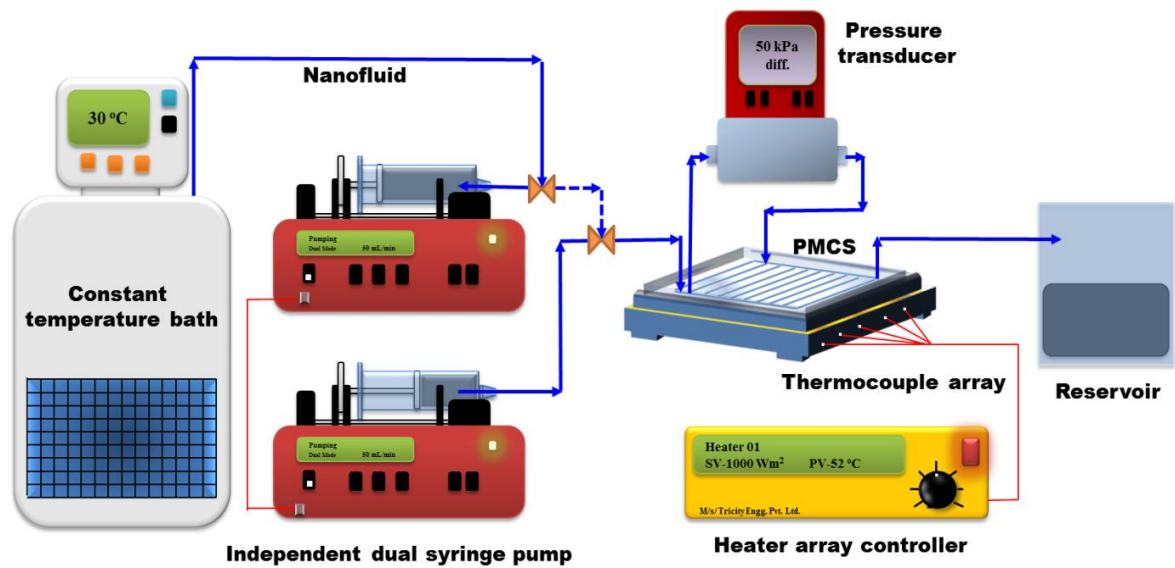

**Figure 3:** **(a)** Microscope image of the surface roughness of the manifold and channels. **(b)** Region of the channel and manifold considered for determining roughness **(c)** Surface roughness profile and average roughness value within channel **(d)** View of the assembled test section.

Distilled water has been used as the working base fluid in the present experiments. The study also involves use of various nanofluids involving nanoparticles which have been procured. The nanoparticles have been characterized for morphology using High resolution

Scanning Electron Microscope (HRSEM) and the same illustrated in fig. 4. Fig 4 (a) illustrates aluminium oxide nanoparticles (Alfa Aeser, 45 nm, spherical), fig. 4 (b) illustrates copper oxide (Nanoshel, 30 nm, oblate), fig. 4 (c) illustrates carbon nanotubes (Alfa Aeser), fig. 4 (d) illustrates graphene (Sisco Research Labs, 3-5 layers) and fig. 4 (e) illustrates silicon oxide (Sigma Aldrich, 14 nm, spherical). The nanofluids have been synthesized by dispersing the required amount of nanoparticles in the base fluid and ultrasonicating (Oscar Ultrasonics, India) the sample with a probe sonicator to stabilize. In case of CNT and graphene, a minute amount of surfactant (Sodium dodecyl sulphate) has been used to stabilize the hydrophobic carbon nanomaterials. Fig. 4 (f) illustrates some representative nanofluids and the nanofluids have been found to be stable for 3-5 days, which is much larger than the experimental time frame. However, since circulatory water bath is used to house the fluid samples before pumping into the test section, further stability of the nanofluids is ensured by the stirring motion of the water bath. The details of the experimental cases considered for the present studies have been tabulated in Table 1 and the associated uncertainties in experimental parameters have been tabulated in Table 2.

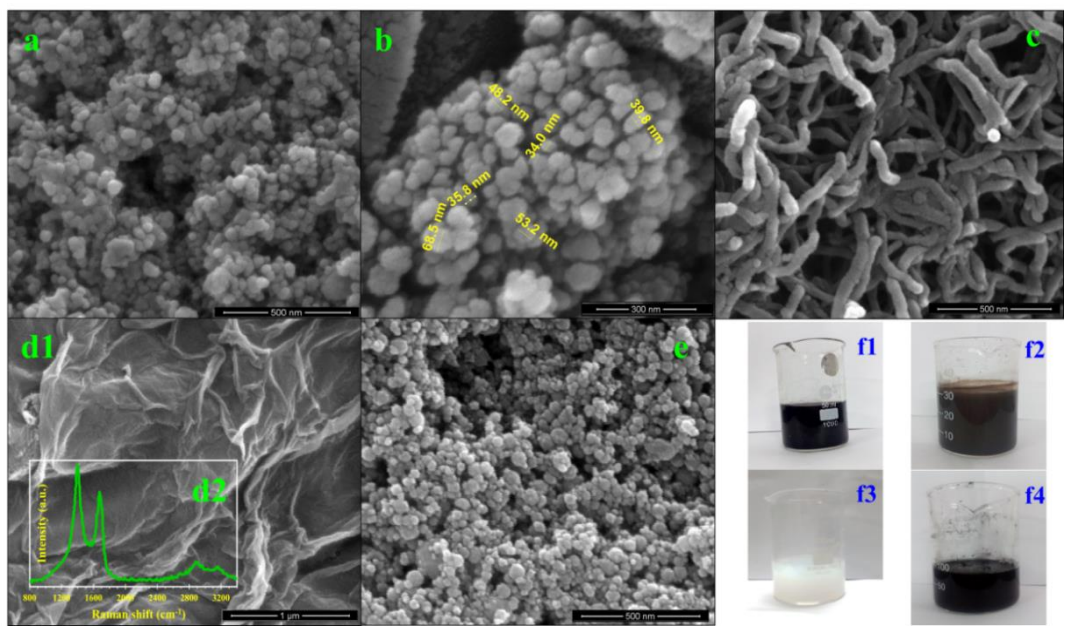

**Figure 4:** Scanning Electron Microscope images of employed nanoparticles **(a)** Aluminium oxide **(b)** Copper oxide **(c)** Carbon nanotubes **(d1)** Graphene and **(e)** Silicon dioxide. **(d2)** illustrates the Raman spectra of the graphene sample. The nanofluid samples employed are **(f1)** graphene **(f2)** copper oxide **(f3)** aluminium oxide and **(f4)** carbon nanotubes.

## 2.2. Simulation details

To model the flow dynamics and heat transfer characteristics of nanofluid in PMCS, a 3-D geometry of the heat spreader with channels embedded has been developed, meshed and solved using the CFD solver Fluent 14.5. The modelling has been done by employing two different mathematical modelling approaches; the Discrete Phase Model (DPM) and the Effective Property Model (EPM). The former one considers the nanofluid as a twin component, non–homogeneous fluid i.e., the dispersed nanoparticle phase is transported by the continuous fluid phase. The latter considers the nanofluid as a single component homogeneous fluid with effective physical properties which are linear functions of fluid and particle material properties determine by weighted averaging with respect to the particle concentration. The flow is considered steady, laminar, and incompressible and the properties are independent of temperature and pressure. A prescribed velocity magnitude corresponding to the corresponding experimental Reynolds number is given at the entry of the inlet manifold. The outlet of the system is exposed to ambient, atmospheric pressure. Uniform heat flux determined from experimental conditions is applied at the bottom and side walls and the top wall is considered as adiabatic for heat transfer cases (a valid assumption as the experimental test section is closed from top using insulating acrylic sheet). The grid independence test has been performed for the considered geometry and optimal grid size has been chosen. The details of numerical simulation cases which have been performed to validate with experimental results have been tabulated in the Table 1. The complete details of the numerical methodology and grid independence test are similar to reports by the present authors [21].

**Table 1:** Details of experimental cases

| Sl. No. | Hydraulic diameter (µm) | Number of channels | $A_C/A_P$ | Re | Configuration |
|---|---|---|---|---|---|
| 1 | 100, 200 | 7 | 0.2 | 10-150 | U, I, Z |
| 2 | 100, 200 | 10 | 0.2 | 10-150 | U, I, Z |
| 3 | 100, 200 | 12 | 0.2 | 10-150 | U, I, Z |

**Table 2:** Associated uncertainties (in %) in experimental parameters

| Sl. No. | Parameter | 100 μm | 200 μm |
|---|---|---|---|
| 1 | $D_H$ | 3.60 | 1.65 |
| 2 | Re | 3.08 | 2.48 |

## 3. Results and discussions

### 3.1. Hydrodynamic performance

Designing of effective PMCS to dissipate high heat fluxes are not only restricted by the objective of high heat removal rates but also to maintain uniformity of temperature within the device. Developing so called thermally efficient (uniform cooling) cooling systems always demand uniform distribution of the working fluid among the conduits. The extent of uniformity in cooling severely depends on the uniformity of distribution of working fluid among the channels of PMCS. Therefore a detailed study of distribution of working fluid among parallel channels is required to understand the thermal performance of any heat exchanger and in the present study it happens to be PMCS. Pressure drop characteristics of water, both predicted by simulations and obtained from experiments, have been illustrated in fig. 5. Fig. 5 (a) illustrates the pressure drop within channels of different configurations viz. U, I and Z at Re=50 using water as working fluid. The computational results have been obtained by solving 3D Navier-Stokes equations within flow domain. From the figure it can be observed that there is a good match (within the error limit of ±10%) between experimental observations and numerical predictions. There is a certain mismatch observed between experimental results and numerical predictions at the first channel of U configuration. In case of experiments, since the flow splits suddenly into first channel from the manifold through a perpendicular turn, it creates lot of flow disturbances at the inlet of first channel. Such stray disturbances are essentially absent in case of computations, and hence higher pressure drop is observed in experiments. However, such pronounced mismatch between experimental and numerical results can be observed near the middle channel of I configurations. In case of I configuration, flow splits near the middle channels, and hence large disturbance and instabilities are experienced at the middle channels which leads to higher pressure drop that is not completely predicted by numerical results.

In case of Z configuration, such peculiar behaviour is not observed. Here, although split of flow occurs at first channel, the flow geometry leads to maximum flow rates in the end channels also. Therefore disturbances are not as pronounced as in case of U and I configurations. This is possible due to the arrangement of channels with respect to manifold such that there are no such major disturbances present at the zones of splitting of the flow field. In case of Z configuration the deviation between experimental and numerical results within 7–8 % error limit has been observed. The uniformity of flow distribution can be gauged quantitatively using the Flow Maldistribution Factor (FMF) [21]. Experimental and numerical results of FMF for three different configurations (U, I and Z) at different Re (10, 50 and 150) have been illustrated in fig. 5(b). From the figure it can be observed that there is a good match between the trends in experimental and numerical results in predicting FMF of three configurations using water as working fluid. FMF of U configuration is highest followed by I and Z configurations. Such observations are due to the highly non uniform distribution of working fluid (which is termed as maldistribution [21, 22]) among the channels which are essentially due to the apparent positioning of channels with respect to manifold. However, as the pressure drop values observed in experiments agree within a margin of ±10%, the ratio of the maximum and minimum pressure drops leads to deviated values of the FMF. The trends of FMF as a function of Re are well matched with numerical predictions. Although flow maldistribution provides insight onto fluid flow and domain interactions, in case of the present study, where thermal management by advective transport is of prime importance, solely the study of flow maldistribution in adiabatic conditions poses very little importance. Flow distribution in presence if heat transfer processes forms the crux of the design parameter for such cooling systems. As viscosity of fluids is a strong function of temperature and maldistribution is strong function of viscosity, flow maldistribution during heat transfer provides more insight on actual design and performance of such PMCS.

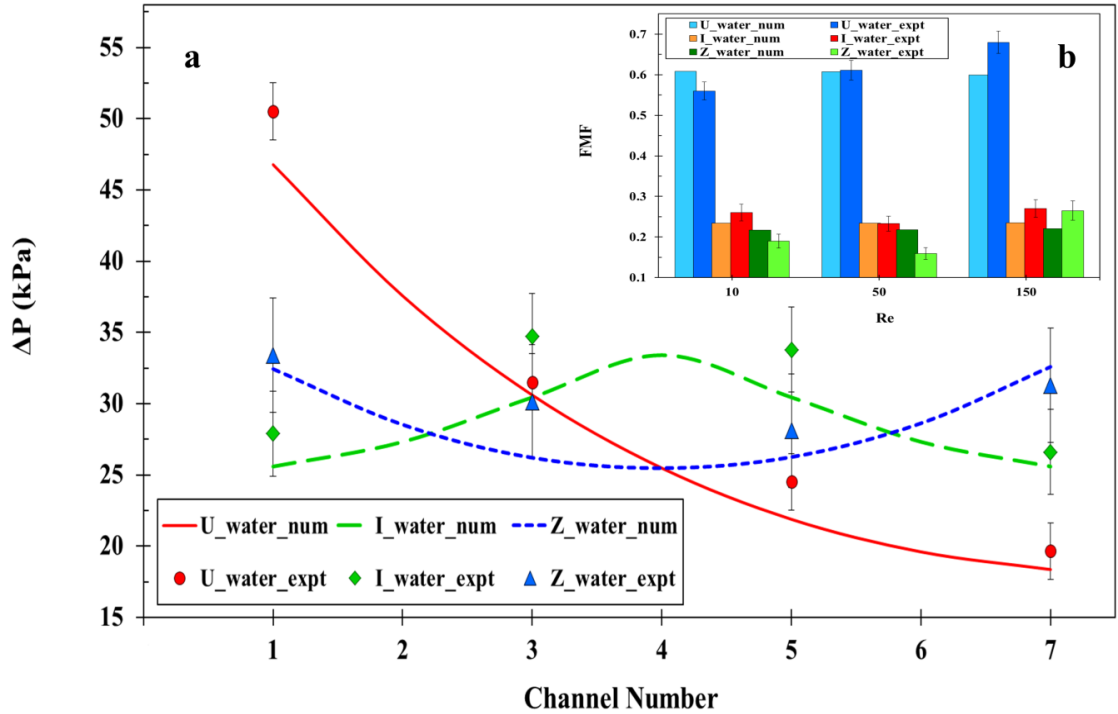

**Figure 5: (a)** Comparison of experimental pressure drop characteristics across channels with simulation results [22] for three configurations (U, I and Z) of 7 channels. **(b)** Comparison of experimental flow maldistribution factor with simulation results for three configurations of 7 channels.

Although enhanced cooling can be achieved using PMCS in prospect of dissipating high heat fluxes from electronic devices, the extent of uniformity of cooling achieved using ordinary conventional fluids such as water as coolants is still questionable. Additionally, improvement of some feature (either geometry or flow features) of PMCS is still required to overcome such shortcomings to at least some extent. One of the best preferences is using highly thermally efficient coolants, for example, nanofluids. Having said that, it is also necessary to understand how so called thermally efficient fluids are hydrodynamically efficient. Fig. 6 represents the comparison of experimental and numerical results of hydrodynamic characteristics of nanofluids compared to water in PMCS. Fig. 6(a) illustrates pressure drop of working fluids within channels at Re=10 and Re=50. Local pressure drop data gives a clear picture of extent of uniformity of flow distribution among the parallel channels. Two numerical models have been employed viz. an Eulerian–Lagrangian twin component Discrete Phase Model (DPM) and property evaluated Effective Property Model (EPM) to model hydrodynamic aspects of nanofluids in parallel microchannels. From the figure it can be observed that modelling of nanofluid in microchannels using EPM formulation over predicts the pressure drop within channels. On the contrary, the DPM formulation predicts relatively less pressure drop and this closely matches with the experimental observations. The EPM considers nanofluids as a single phase fluid with effective physical properties estimated based on weighted average method using the particle concentration as the weighing parameter. The DPM models nanofluid as a two phase non homogeneous fluid (fluid + solid) and considers all predominant diffusion and migration effects of the nanoparticles in the base fluid, such as Brownian motion, thermophoresis, Saffman lift, drag, etc. DPM formulation has been reported to be an efficient and suitable tool to model nanofluid flows in microchannels [21-23]. From the figure it is evident that there are no appreciable differences between pressure drops of water and nanofluid as working fluids. Hence, nanofluids can be used as working fluids to address cooling challenges of high heat flux generation devices without suffering large increments in the pumping requirements. Fig. 6(b) has been plotted to represent pressure drop of alumina water nanofluid within PMCS at different concentrations for both DPM predictions and experimental data. Experimental pressure drop data is observed to match well with the numerical results with 10-12% error. The source of the error is possibly due to the enhanced friction factor caused by the channel roughness (imparted by the micro–machining process) while the numerical domain is completely smooth and the frictional pressure drop predicted is lesser.

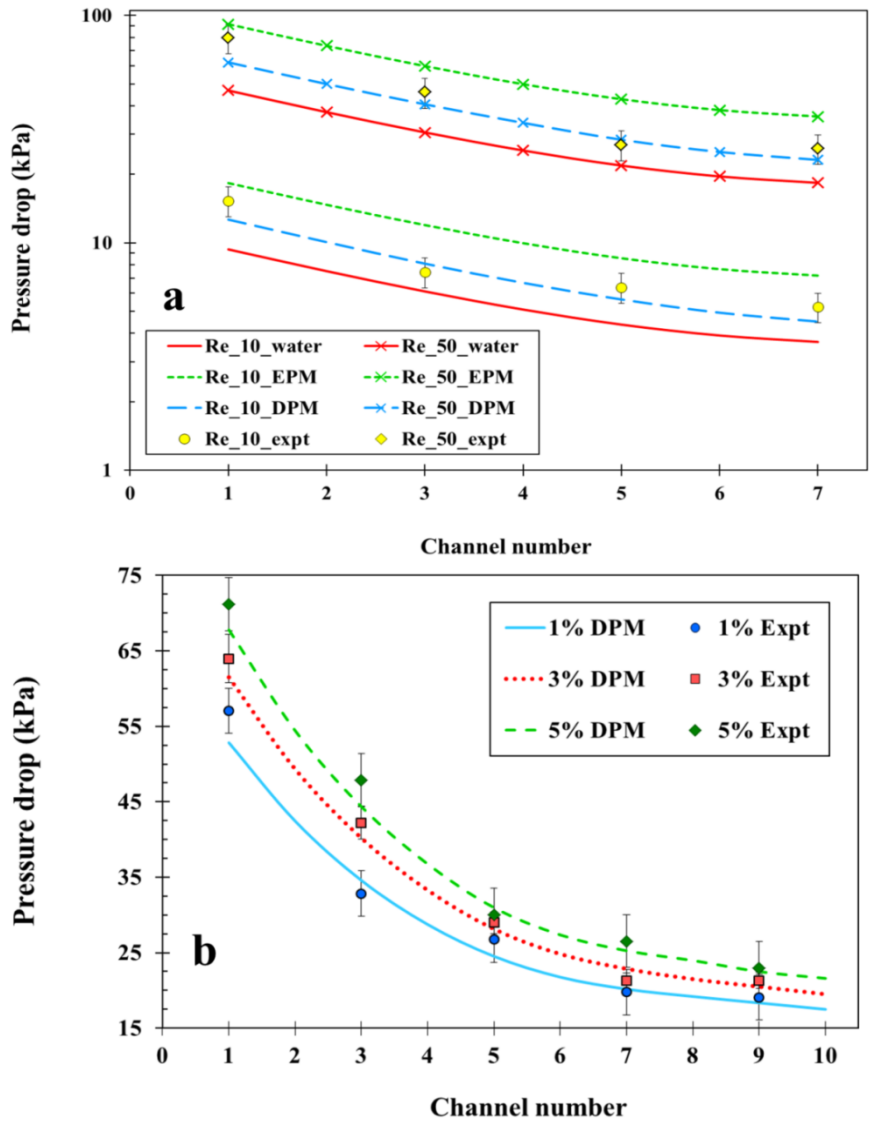

**Figure 6: (a)** Comparison of experimental and numerical Pressure drop (numerical using both DPM and EPM formulation) characteristics with respect to channel number for U configuration. Pressure drop for flow of water and 5 wt. % alumina-water nanofluid at Re=10 &Re=50. **(b)** Experimental and numerical pressure drop characteristics with respect to channel number at different nanofluid concentrations (1, 3 and 5 wt. % Alumina) at Re 50.

Although there is a marginal increase in pressure drop by adding nanoparticles to base fluid; study of flow characteristics when diameter of the channel increases is required as an acute knowledge of the pressure characteristics is important for design of the microscale pumping systems suitable for such miniaturized flow devices. According to Kandlikar [25] classification of channel hydraulic diameter, any channel with hydraulic diameter between 1-200 μm is said to be microchannel. Pressure drop characteristics of flow in two different hydraulic diameter microchannels have been plotted in Fig. 7. Since the major migration phenomena like Brownian and thermophoretic motions are in the order of nanoscale, the study of effect of size of flow path on the extent of exploiting the migration phenomena in diffusion of heat is very important. Fig. 7(a) has been plotted to understand the hydrodynamics (which is prerequisite to understand heat transfer) of working fluid in two different hydraulic diameter PMCS. The figure has been plotted for pressure drop comparison of numerical (DPM) and experimental results of 100 and 200 μm hydraulic diameters with nanofluid as working fluid. From the pressure drop plot, it is observed that there is a relatively uniform distribution of fluid in case of 100 μm compared with 200 μm hydraulic diameter. It is expected due to high pressure drop offered among channels which leads to more uniform distribution of fluid in case of 100 μm compared to 200 μm hydraulic diameter [9]. The effect of hydraulic diameter on cooling performance of PMCS has been discussed in further sections. Flow characteristics of nanofluid of different concentrations in PMCS with I and Z configurations at different Re have been illustrated quantitatively (using the FMF) in Fig. 7(b). It can be observed that there is a good match between the trends of experimental results and numerical predictions but the error between both the predictions is quite high, ~9-11 %. FMF of experimental results is more, implying poor distribution of fluid among channels in case of experiments and it is possible be due to roughness and error in geometry caused due to fabrication limitations which is absent in case of numerical modelling. It can be inferred from the figure that in case of I and Z configurations as Re increases maldistribution of coolant among channels reduces by a small extent. Z configuration shows least maldistribution followed by I and it is extreme for U in case of water as well as nanofluid as coolants. Hence Z configuration PMCS employing nanofluid as coolant can be used as an effective cooling technique to cool electronic devices, under the constraint that '*the heat produced by device is uniform*'. The thermal performance of U, I and Z configuration PMCS has been analysed in further sections.

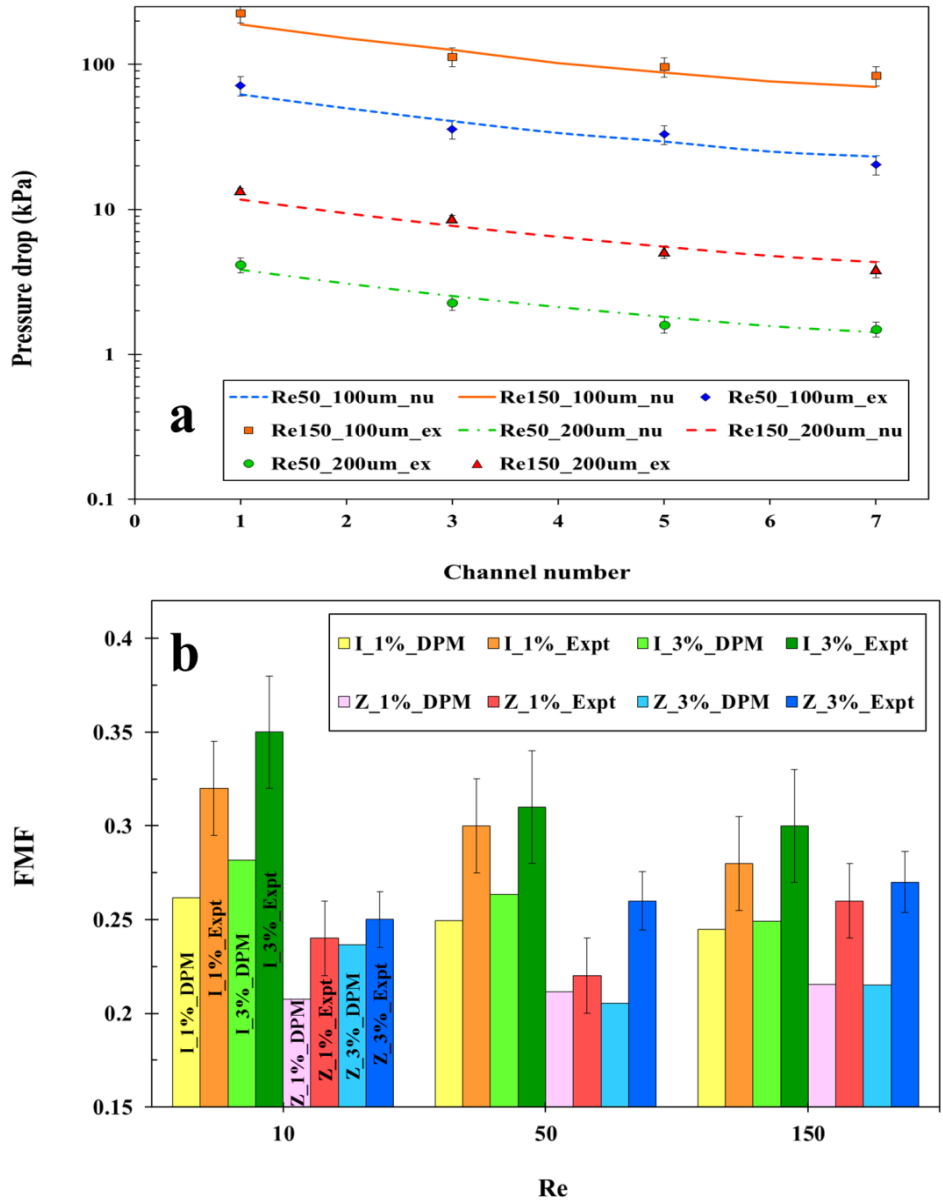

**Figure 7: (a)** Comparison of pressure drop characteristics within channels of U configuration for two different hydraulic diameters (100 & 200 μm) at Re=50 & Re=150. Working fluid used is alumina (Al$_2$O$_3$)-water nanofluid of 5 wt. % concentration. **(b)** Comparison of Flow Maldistribution Factor (FMF) of I and Z configurations using Alumina-water nanofluid of three concentrations (1, 3 & 5 wt. %) as working fluids.

Employing nanofluids as coolants is a good option to achieve uniform cooling of microelectronic devices and modules thereby making the analysis of the performance of variant nanofluids as coolants important. Since nanofluids are reported to act as smart fluids in micro flow paths [21, 22], it is essential to know the hydro dynamic behaviour of the same for design of the pumping system. Comparison of hydrodynamic characteristics of different types of nanofluids in PMCS has been illustrated in fig. 8. Five types of nanofluids have been used for the study by dispersing aluminium oxide, copper oxide, silicon dioxide, CNT and graphene nanoparticles in water. Pressure drop of nanofluids across the first channel of U configuration PMCS having 7 channels have been compared with water at Re 50. In hydrodynamics point of view alumina, CNT and graphene have approximately equal pressure drop characteristics when compared with water. But the scenario might be different when it comes to heat transfer perspective. Since heat transport by particles is mostly depends on morphology [26] and thermal properties of nanoparticles, different nanofluids possess different thermal performance even though the hydrodynamic characteristics are same. The pressure drop exhibited by copper oxide nanofluid is higher whereas silicon dioxide nanofluid is less compared with other nanofluids considered. It is expected because of high density of CuO-water nanofluid which experiences high flow resistance consequently produces high pressure drop. Likewise, the low density of silicon oxide and high affinity for water leads to nominal increment in viscosity and leads to low pressure drop. Hence, CuO is not recommended for the high pressure drop and silicon oxide for the poor thermal characteristics, leaving aluminium oxide, CNT and graphene based nanofluids as choicest for thermal mitigation.

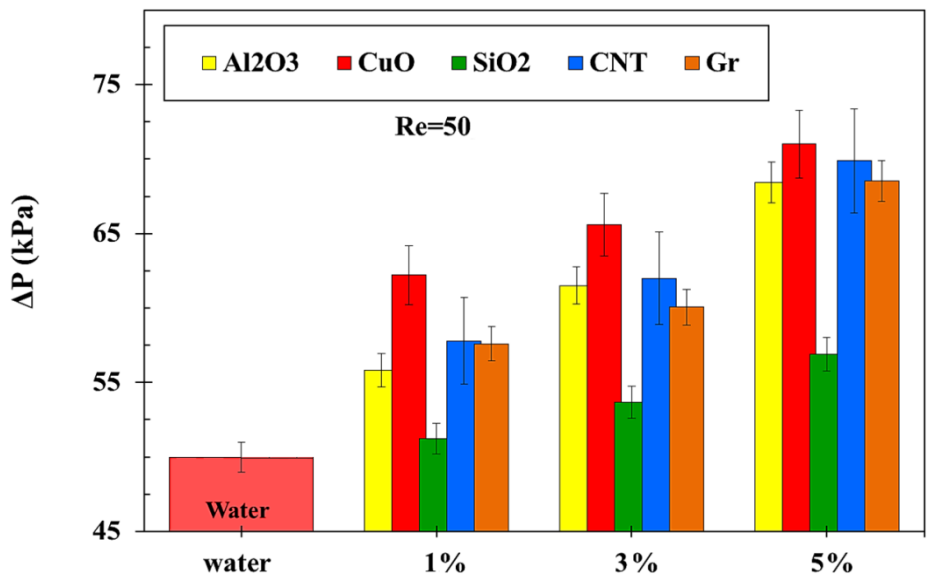

**Figure 8:** Comparison of pressure drop of water with respect to different nanofluids in a first channel among 7 channels of U configuration at Re=50. Nanofluids used are $Al_2O_3$, CuO, $SiO_2$, CNT, Gr of concentrations 1, 3 & 5 wt. %.

## 3.2. Thermal performance

Having discussed the hydrodynamic characteristics of PMCS with respect to working fluids and geometrical conditions, it is now essential to analyse the effect of hydrodynamic characteristics on the thermal performance of PMCS. The utmost objective of designing PMCS is to reduce the maximum temperature produced within the device since high chip temperature leads to failure of electronic devices. The main thrust in the design of cooling systems (PMCS) for any microelectronic device is the alliance of (a) reducing maximum temperature rise within the device and (b) to improve the uniformity of cooling. Fig. 9 has been plotted for maximum temperature and average temperature (which represents the uniformity of cooling) produced in PMCS at different applied uniform heat fluxes (the applied heat flux is nothing but heat load to PMCS from the associated electronic device). The working fluid used is water at flow Re of 10 and 50. The figure has been plotted for both experimental results and numerical predictions.

As discussed earlier, for a given configuration, thermal performance of PMCS tremendously depends on flow distribution among channels and the same can be observed from the figure. Fig. 9(a) illustrates the maximum temperature and the average temperature generated in U configuration at different heat fluxes using water as heat transfer fluid. Good agreement has been observed between experimental results and numerical predictions (DPM modelling). Since U configuration is known to be highly a flow maldistributed case, the maximum temperature and average temperature within system are always higher compared to I and Z cases. Comparison of maximum temperature in case of U, I and Z configurations has been illustrated in fig. 9 (b). It is observed from the figure that maximum temperature is highest in case of U compared to I and Z (number of channels, area ratio of channel to manifold and hydraulic diameter of channel kept same for all three configurations to bring out the effect of configuration alone on thermal performance of PMCS) and the poor hydrodynamic characteristics of U configuration is solely responsible for its poor thermal performance.

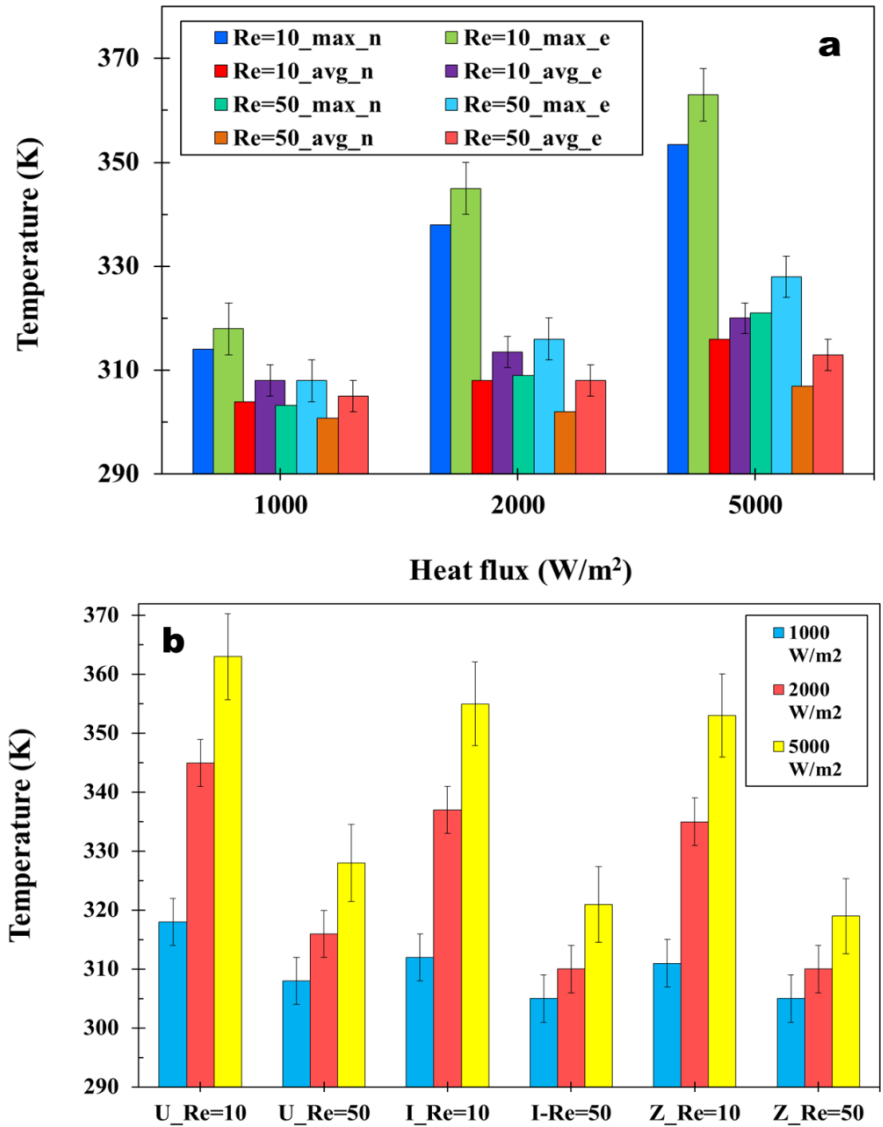

**Figure 9: (a)** Comparison of experimental and numerical results of maximum temperature and average temperature produced in U configuration test section at different heat fluxes using water as working fluid at Re=10 & Re=50. **(b)** Comparison of maximum temperature raise within U, I and Z configurations for different heat fluxes at Re=50 and water as working fluid. The hydraulic diameter of the channels is 100 μm.

The basic utility of PMCS is to mitigate overshoot of temperatures at the hot spot and to achieve uniform cooling throughout the PMCS. In reality, single phase fluids such as water are not expected to fulfil the cooling requirements. Employing nanofluids as working fluid is one of the best alternative solutions to attain enhanced cooling as these fluids demonstrate smart features [21, 22] due to particle migration which enhances heat transfer characteristics. Experimental results of maximum and average temperatures within U type PMCS (10 channels) at different Re and at different uniform heat fluxes have been illustrated in fig. 10. From the figure it can be observed that using nanofluid enhances cooling compared to water. There are two possibilities which exist in present case so as to enhance cooling, viz. 1) increasing flow velocity for a given working fluid and 2) increasing the thermal conductivity as well as convective heat transfer capabilities of the fluid at the same flow velocity. The former one leads to increase in heat transfer coefficient, however, at the cost of enhanced pressure drop across the flow domain. The latter one increases the thermal conductivity of the base fluid as well as increases the convective transport coefficient much higher than that obtained by increasing Re by a certain extent [28]. Increasing flow Re along with adding high thermal conductivity particles in the base fluid will serve the purpose of achieving uniform cooling. However due to high pressure drops, increase in Re number is not an appreciable choice in case of micro flow paths. Hence using nanofluids with moderate Re values is an intelligent choice to achieve the necessary and sufficient objectives for a PMCS employed to cool a microprocessor. The decrease in average temperature within the PMCS as well as the peak temperature by employing nanofluids (as illustrated in fig. 10) essentially supports the fact that nanofluids are promising coolants for both mitigating hot spots as well as overall uniformity in cooling.

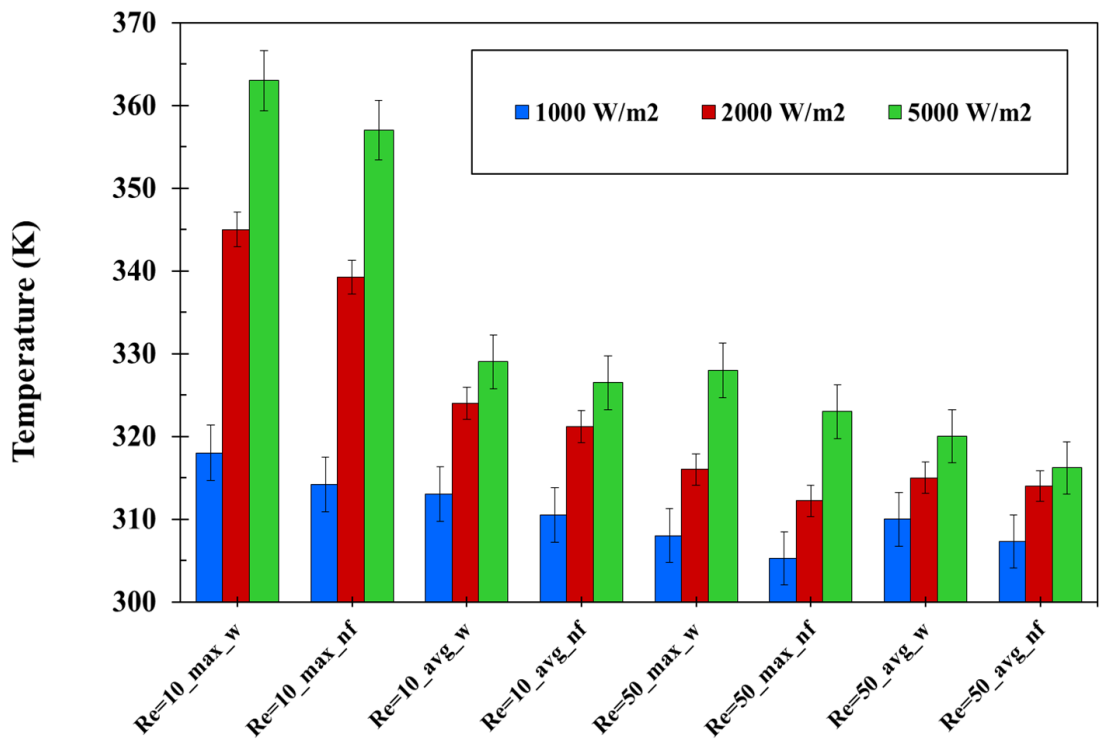

**Figure 10:** Experimental results of maximum temperature and average temperature produced in U configuration PMCS at different heat fluxes using water and alumina-water nanofluid (3 wt. %) as working fluids at Re=10 & Re=50.

Thermal performance of U, I and Z configuration PMCS have been compared using different nanofluids as working fluids in figure 11. Water based aluminium oxide, copper oxide, multi-walled carbon nanotubes (MWCNT) and graphene nanofluids have been employed as working fluids and compared with the performance of water. From the figure it can be seen that the cooling capability of CNT and graphene based nanofluids is much pronounced compared with alumina and copper oxide based nanofluids. As illustrated in fig. 8, though alumina, CNT and graphene nanofluids possess similar pressure drop characteristics in microchannels but the scenario is different when it comes to heat transfer. Since heat transport by particles mostly depends on thermal properties and morphology of nanoparticles, different nanofluids possess different thermal performance even though hydrodynamic characteristics are same. From thermal conductivity perspective, MWCNT and graphene which are highly thermally conductive materials are expected to have high cooling capability when in the dispersed form. MWCNT and graphene are excellent thermal conductors along the tubes and along the flake faces, exhibiting a property known as ballistic conduction of the heat carrying phonons, but good insulators lateral to the tube axis where the phonon conduction is largely restricted. Hence, thermal performance of MWCNT nanofluids and graphene nanofluid essentially depends on the orientation of tubes and flakes in the flow field [27]. Though available correlations (effective property based) predict higher effective thermal conductivity, but the actual thermal performance of such special heat transfer fluids (MWCNT and graphene) are noticeably less than expected from theoretical predictions. However, in spite of this, the thermal characteristics of such nanofluids are high and CNT and graphene nanofluids show considerably high cooling performance compared with base fluid (water), $Al_2O_3$ and CuO nanofluids. At applied uniform heat flux of 5000 $W/m^2$ and Re=50 flow conditions for a PMCS of U configuration using CNT-water nanofluid, the maximum temperature at the hot spot region dropped by 7±1 $^oC$ compared with water as working fluid. Essentially it shows that such carbon nanostructure based nanofluids can be employed as efficient heat transfer fluids in place of base fluids to achieve required objectives to cool MEMS.

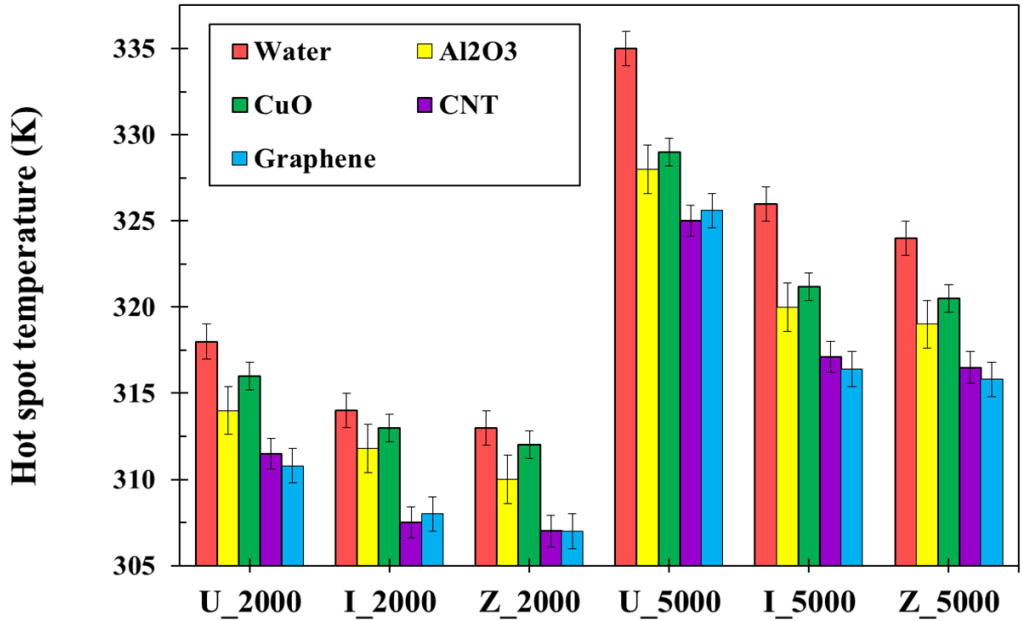

**Figure 11:** Experimental results of hot spot core temperature (maximum temperature) generated in U, I and Z configurations PMCS at 2000W/m$^2$ and 5000W/m$^2$ heat fluxes using water and different types of nanofluids as working fluids with Re=50. Number of channels of PMCS are 10.

In order to validate the pattern of hot spot formed due to the unique flow nature within each configuration is same as that predicted by simulations, qualitative comparison of thermal contours from simulations reported by the present authors [22] with respect to experimental results of PMCS have been represented in fig. 12. The operating conditions of figure are as follows; uniform heat flux applied is 5000W/m$^2$, working fluid is alumina-water nanofluid at Re=50. Thermal images of the experimental test section have been captured using an infrared camera (FLIR E40, thermal sensitivity < 0.1 $^o$C at 30$^o$C). From the figure it can be observed that there is a good qualitative agreement between numerical thermal contours and thermal images of experiments with respect to hot spot position. Thermal images of experiments have been observed to be more diffused in nature compared with numerical contours. In case of experiments, capturing thermal images is solely possible through acrylic cover which is on top of the test section and the temperature profile spreads considerably compared to the profile at the PMCS. Accordingly, the thermal contours visible are that present on the surface of the acrylic cover. Consequently, the maximum temperatures obtained from experimental thermal images are less than numerical predictions, due to the temperature drop caused by the conductive resistance of acrylic cover. However, the images validate that the overall morphology of the hot spot is similar to that predicted.

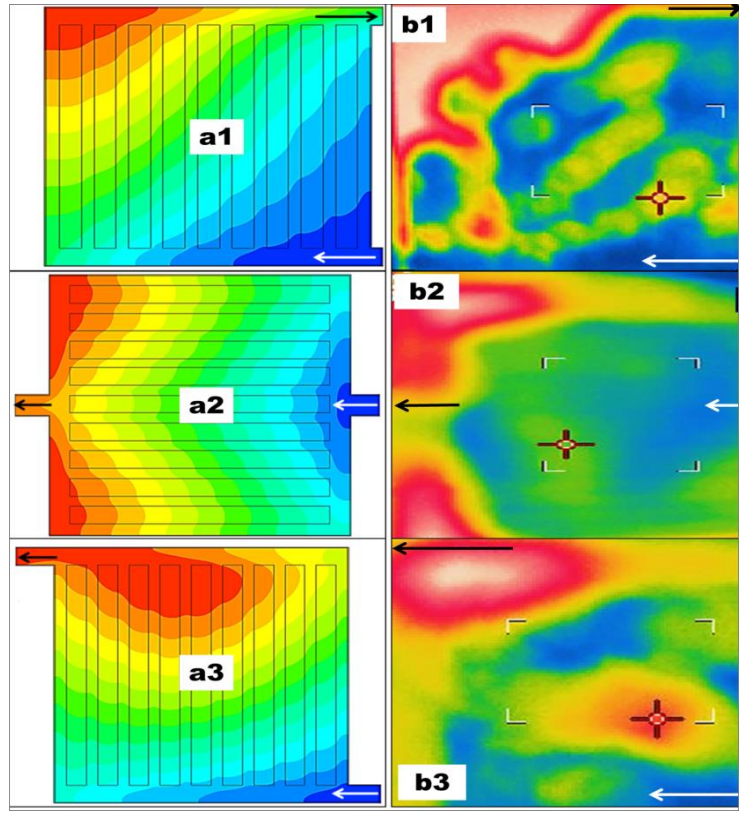

**Figure 12:** Comparison of (a) numerical results of thermal contours of U, I and Z configurations with (b) thermal images of experiments using thermal camera. 1, 2 and 3 represent U, I and Z cases respectively.

Detailed fundamental parametric studies (involving the effects of flow configuration, Re, number of channels, channel hydraulic diameter and manifold area) on the extent of flow maldistribution has been well established in reported literature [9, 14]. But detailed studies of the effects of flow maldistribution patterns on the thermal performance of PMCS have not been established. Accordingly, the present article concentrates on the applicability point of view i.e. to choose the best suitable PMCS configuration to efficiently cool electronic devices (effect of flow configuration induced flow maldistribution on thermal performance of PMCS). Furthermore, the objective is to also bring out the smart effects of nanofluids as efficient thermal performance fluids in PMCS. To show the capability of microscale heat transfer as potential cooling solution, the dependency of thermal performance of PMCS on channel hydraulic diameter and number of channels have been shown in the fig. 13. As discussed earlier, according to Kandlikar's [25] classification, any channel with hydraulic diameter between 1-200 μm can be termed as a microchannel. When nanofluid is employed as a working fluid, there exists particle migration phenomenon. The mean free path of such migratory diffusion is of the order of ~ 100 nm (i.e. order of particle diameter) and hence leads to enhanced transport of heat within microscale flow domains (i.e. where the channel hydraulic diameter is at best 1-3 orders of magnitude larger than the particle diffusion length). From the figure it can be observed that the thermal performance of PMCS reduces with increment in the channel hydraulic diameter and all other factor remaining unchanged. An increment of ~6 $^0$C in the maximum temperature is observed within domain when water is employed as working fluid for the larger diameter channel. The effect of number of channels on the thermal performance of PMCS has been illustrated in fig. 13 (b). From the figure it can be observed that as the number of parallel channels of PMCS increases (all other geometrical and flow conditions being constant) thermal performance of the PMCS decreases. Such a phenomenon occurs because in case of U configuration, as the number of channels increases, the flow maldistribution increases and consequently thermal performance of PMCS decreases. However, it is noteworthy that increment in number of channels would lead to the intuition that effective heat transfer would increase due to increased area of heat transfer. Such a phenomenon is only possible when the port area is increased to allow better distribution of fluid within the channels. In the present case the ratio of port area to channel area has been held constant (as the objective is to only see the effect of number of channels and the fact that increasing port area is not justifiable due to space constraints in real time systems). Hence, in spite of increased heat transfer area the maldistribution of fluid increases with increase in channel number and the thermal performance drops.

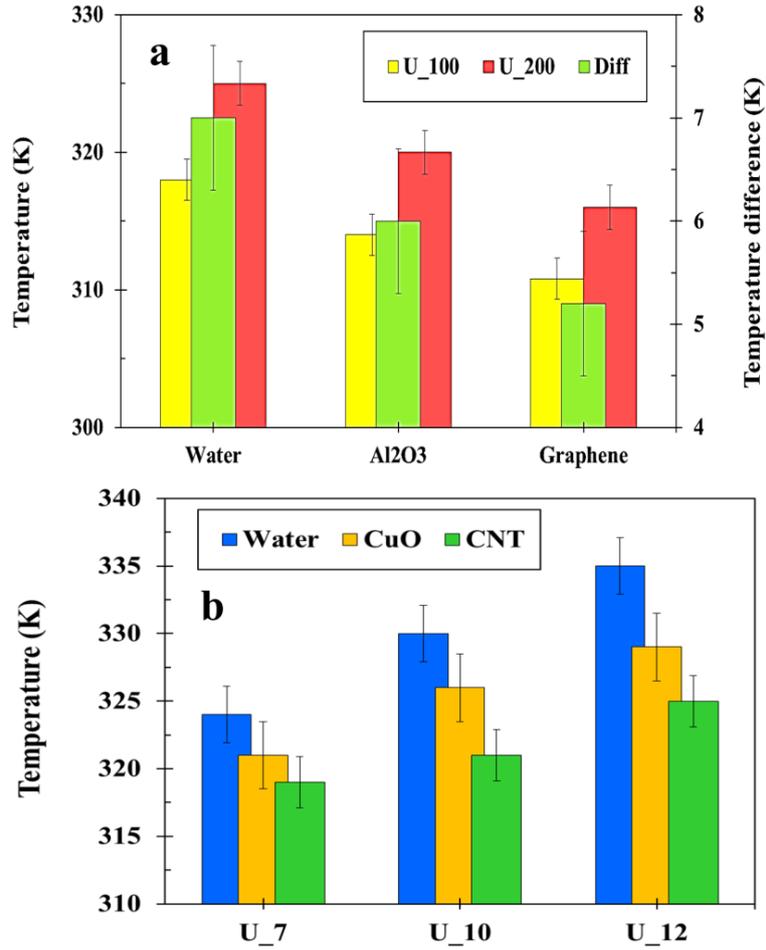

**Figure 13:** Experimental results of maximum temperature generated within the U configuration PMCS. (a) Maximum temperature generated for three different working fluids and with two different channel hydraulic diameters. (b) Maximum generated temperature for different number of channels using three different working fluids.

To quantify the two important aspects discussed in preceding sections, i.e. 1) selection of the best configuration for uniform cooling of a given geometrical condition and for an applied heat load and 2) increasing heat transfer performance in a given configuration employing nanofluid compared with base fluid, a parameter termed as *Figure of Merit* (FoM) has been proposed [22]. FoM is defined in a way that it simultaneously considers enhancing the uniformity of cooling as well as reducing the maximum temperature within the PMCS. Accordingly, a higher value of the FoM establishes a better cooling solution for the system at hand. FoM can be mathematically expressed as $\xi = \frac{1}{(T_{max}-T_{min})(T_{max}-T_{inlet})}$ where term $(T_{max}-T_{min})$ takes care of uniformity of cooling and the term $(T_{max}-T_{inlet})$ take care of maximum temperature rise in the domain. To achieve uniformity of cooling the difference between maximum and minimum temperatures within domain should be least and to attain reduced peak temperatures, difference between the maximum and inlet temperatures should also be minimal. The two terms being in the denominator, higher FoM value denotes a system as highly thermally efficient system. Fig. 14 has been plotted for experimental results and numerical predictions of FoM for all three configurations. Water and alumina-water nanofluid of 5 wt. % have been used as working fluids and flow conditions are Re=100. From the figure it can be observed that FoM of Z configuration employing nanofluid as working fluid is having the highest FoM value compared with other possible combinations when applied heat flux is uniform and hence the configuration–nanofluid pair is the best suited design for efficient thermal management of microprocessors via the present concept. Similarly, due to the fact that the U configuration, due to its flow pattern induces a hot spot with high temperature, it exhibits least FoM. In fact, even upon usage of nanofluids, the hot spot is not sufficiently cooled and hence the increment in FoM with respect to water is much less compared to the I or Z cases.

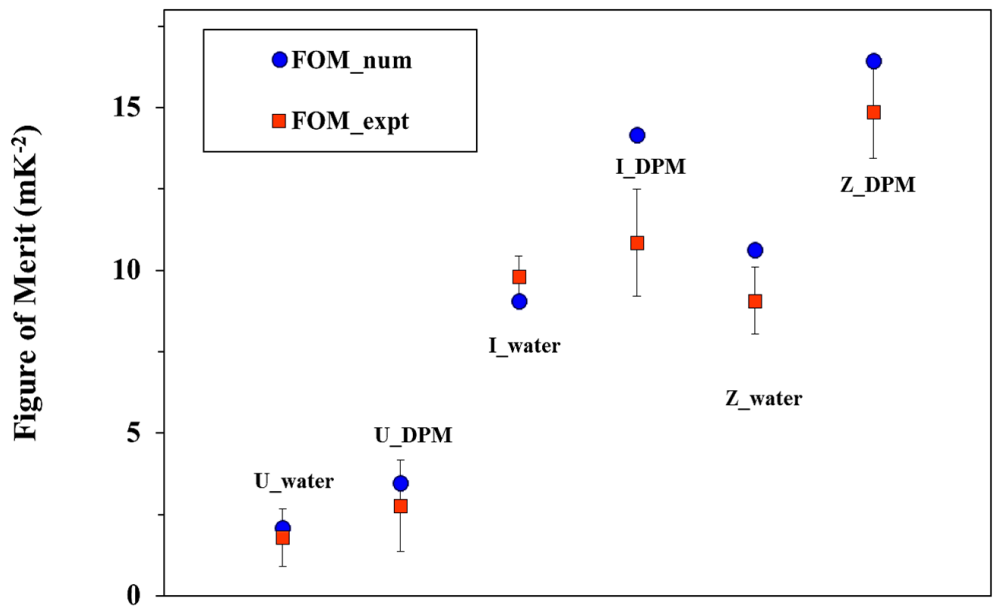

**Figure 14:** Comparison of experimental and numerical results for Figure of Merit (FoM) plotted for three configurations (U, I and Z) for operating conditions are Re=100 and alumina (5 wt. %)-water nanofluid and water as working fluids.

## 4. Conclusions

To infer, the present article discusses a novel methodology for near-active cooling of MEMS using smart fluids and PMCS. The article discusses specific design of the PMCS such that it can be implemented at ease on the heat spreader of a modern microprocessor to obtain near-active cooling. Comprehensive experimental and numerical studies have been carried out to understand the effects of three different flow configurations (U, I and Z) of PMCS have been discussed. The study not only focusses on the thermofluidics due to flow configuration, nanofluids of aluminium oxide, copper oxide, silicon oxide, CNT and graphene have also been employed to achieve the desired mitigation of hot spot temperatures and to improve the uniformity of cooling. Two numerical modelling methods, DPM and EPM have been used to model nanofluids as working fluid and DPM predictions have been observed to match well with experiments. In order to quantify the cooling performance of a particular configuration–fluid pair, a Figure of Merit (FoM) has been proposed from thermodynamics point of view. From the FoM It has been observed that the Z configuration employing nanofluid is the best suitable solution for uniform thermal loads to achieve uniform cooling as well as reducing maximum temperature. The results are found to be very promising and a feasible approach for thermal management of microprocessor systems.

## Acknowledgements

LSM and PD would like to thank IIT Ropar for the partial financial support for the present work and also Dr. H. Tyagi of Department of Mechanical Engineering, IIT Ropar for the thermal camera. LSM and PD would also like to thank Dr. A. Agrawal of Department of Mechanical Engineering for the micromachining facility. LSM and PD also thank Mr. Varinder (Machinist) and Mr. Rakesh (Technical Assistant, IIT Ropar) for CNC machining facilities, Mr. Ramkumar (Technical Assistant, IIT Ropar) for help with fabrication of heater assembly and Mr. AR Harikrishnan (Scholar, IIT Madras) for technical discussions. LSM would also like to thank the Ministry of Human Resource Development, Govt. of India for the doctoral scholarship.

# References

1. D. B. Tuckerman and R. F. W. Pease, High performance heat sinking for VLSI, IEEE Electron Device Letters 2 (5) (1981) 126-129.
2. W. Qu, and Mudawar. I, Analysis of three-dimensional heat transfer in micro-channel heat sinks, International Journal of heat and mass transfer 45 (19) (2002) 3973-3985.
3. X. F. Peng, G. P. Peterson, and B. X. Wang, Heat transfer characteristics of water flowing through microchannels, Experimental Heat Transfer An International Journal 7 (4) (1994) 265-283.
4. Q. Weilin, G. M. Mala and L. Dongqing, Pressure-driven water flows in trapezoidal silicon microchannels, International Journal of Heat and Mass Transfer, 43 (3) (2000) 353-364.
5. B. Xu, K. T. Ooti, N. T. Wong and W. K. Choi, Experimental investigation of flow friction for liquid flow in microchannels, International Communications in Heat and Mass Transfer, 27 (8) (2000) 1165-1176.
6. Z. Y. Guo and Z. X. Li, Size effect on single-phase channel flow and heat transfer at microscale, International Journal of Heat and Fluid Flow, 24 (3) (2003) 284-298.
7. S. G. Kandlikar, History, advances, and challenges in liquid flow and flow boiling heat transfer in microchannels: a critical review, Journal of Heat Transfer, 134 (3) (2012) 034001.
8. S. G. Kandlikar, High flux heat removal with microchannels—a roadmap of challenges and opportunities, Heat Transfer Engineering, 26 (8) (2005) 5-14.
9. V. M. Siva, A. Pattamatta and S. K. Das, Investigation on flow maldistribution in parallel microchannel systems for integrated microelectronic device cooling, IEEE Transactions on Components, Packaging and Manufacturing Technology, 4 (3) (2014) 438-450.
10. G. Hetsroni, A. Mosyak and Z. Segal, Nonuniform temperature distribution in electronic devices cooled by flow in parallel microchannels, IEEE Transactions on Components and Packaging Technologies, 24 (1) (2001) 16-23.
11. K. K. Nielsen, K. Engelbrecht, , D. V. Christensen, J. B. Jensen, A. Smith and C. R. H. Bahl, Degradation of the performance of microchannel heat exchangers due to flow maldistribution, Applied Thermal Engineering, 40 (2012) 236-247.
12. V. M. Siva, A. Pattamatta and S. K. Das, Effect of flow maldistribution on the thermal performance of parallel microchannel cooling systems, International Journal of Heat and Mass Transfer, 73 (2014) 424-428.
13. V. M. Siva, A. Pattamatta and S. K. Das, A numerical study of flow and temperature maldistribution in a parallel microchannel system for heat removal in microelectronic devices, Journal of Thermal Science and Engineering Applications, 5 (4) (2013) 041008.

14. G. Kumaraguruparan, R. M. Kumaran, T. Sornakumar and T. Sundararajan A numerical and experimental investigation of flow maldistribution in a micro-channel heat sink, International Communications in Heat and Mass Transfer, 38 (10) (2011) 1349-1353.
15. M. Pan, D. Zeng, Y. Tang and D. Chen, CFD-based study of velocity distribution among multiple parallel microchannels, Journal of computers, 4 (11) (2009). 1133-1138.
16. C. Anbumeenakshi and M. R. Thansekhar, Experimental investigation of header shape and inlet configuration on flow maldistribution in microchannel, Experimental Thermal and Fluid Science, 75 (2016) 156-161.
17. W. Escher, T. Brunschwiler, N. Shalkevich, A. Shalkevich, T. Burgi, B. Michel and D. Poulikakos, On the cooling of electronics with nanofluids. Journal of heat transfer, 133 (5) (2011) 051401.
18. J. Lee and I. Mudawar, Assessment of the effectiveness of nanofluids for single-phase and two-phase heat transfer in micro-channels, International Journal of Heat and Mass Transfer, 50 (3) (2007) 452-463.
19. M. Nazari, M. Karami and M. Ashouri, Comparing the thermal performance of water, Ethylene Glycol, Alumina and CNT nanofluids in CPU cooling: Experimental study, Experimental Thermal and Fluid Science, 57 (2014) 371-377.
20. S. K. Das, N. Putra, P. Thiesen and W. Roetzel, Temperature dependence of thermal conductivity enhancement for nanofluids, Journal of heat transfer, 125 (4) (2003) 567-574.
21. L. S. Maganti, P. Dhar, T. Sundararajan and S. K. Das, Particle and thermo-hydraulic maldistribution of nanofluids in parallel microchannel systems, Microfluidics and Nanofluidics, 20 (109) (2016),
22. L. S. Maganti, P. Dhar, T. Sundararajan and S. K. Das, Thermally 'smart' characteristics of nanofluids in parallel microchannel systems to mitigate hot spots in MEMS, (2016) (in press) DOI: 10.1109/TCPMT.2016.2619939
23. P. K. Singh, P. V. Harikrishna, T. Sundararajan and S. K. Das, Experimental and numerical investigation into the hydrodynamics of nanofluids in microchannels, Experimental Thermal and Fluid Science, 42 (2012) 174-186.
24. S. Savithiri, P. Dhar, A. Pattamatta and S. K. Das, Particle fluid interactivity deteriorates buoyancy driven thermal transport in nanosuspensions: A multi component lattice Boltzmann approach, Numerical Heat Transfer A, 70 (3) (2016) 260-281.
25. S. G. Kandlikar, History, advances, and challenges in liquid flow and flow boiling heat transfer in microchannels: a critical review, Journal of Heat Transfer, 134 (3) (2012) 034001.
26. P. Dhar, M. H. D. Ansari, S. S. Gupta, V. M. Siva, T. Pradeep, A. Pattamatta & S. K. Das, Percolation network dynamicity and sheet dynamics governed viscous behaviour of polydispersed graphene nanosheet suspensions, Journal of nanoparticle research, 15 (12) (2013) 1-12.

**27.** P. Dhar, S. S. Gupta, S. Chakraborty, A. Pattamatta & S. K. Das, The role of percolation and sheet dynamics during heat conduction in poly-dispersed graphene nanofluids, Applied Physics Letters, 102 (16) (2013) 163114.

**28.** K. B. Anoop, T. Sundararajan & S. K. Das, Effect of particle size on the convective heat transfer in nanofluid in the developing region, International journal of heat and mass transfer, 52 (9) (2009) 2189-2195.